# The Enhancement of Software Delivery Performance through Enterprise DevSecOps and Generative Artificial Intelligence in Chinese Technology Firms.

Jun Cui[1, a, *]

1 Solbridge International School of Business, Ph.D., Daejeon, 34613, Republic of Korea.

a jcui228@student.solbridge.ac.kr

* Correspondence: Jun Cui, jcui228@student.solbridge.ac.kr.

*Abstract*—This study investigates the impact of integrating DevSecOps and Generative Artificial Intelligence (GAI) on software delivery performance within technology firms. Utilizing a qualitative research methodology, the research involved semi-structured interviews with industry practitioners and analysis of case studies from organizations that have successfully implemented these methodologies. The findings reveal significant enhancements in research and development (R&D) efficiency, improved source code management, and heightened software quality and security. The integration of GAI facilitated automation of coding tasks and predictive analytics, while DevSecOps ensured that security measures were embedded throughout the development lifecycle. Despite the promising results, the study identifies gaps related to the generalizability of the findings due to the limited sample size and the qualitative nature of the research. This paper contributes valuable insights into the practical implementation of DevSecOps and GAI, highlighting their potential to transform software delivery processes in technology firms. Future research directions include quantitative assessments of the impact on specific business outcomes and comparative studies across different industries.

## I. Introduction

The increasing complexity of software development, alongside the growing demand for rapid and reliable software delivery, has led organizations to adopt integrated approaches like DevOps and its evolved counterpart, DevSecOps. DevSecOps not only emphasizes collaboration between development and operations teams but also incorporates security practices throughout the software development lifecycle. Concurrently, Generative Artificial Intelligence (GAI) has emerged as a transformative technology, capable of automating various aspects of software development, including coding, testing, and deployment. By leveraging GAI within a DevSecOps framework, technology firms can enhance their software delivery performance, mitigate security vulnerabilities, and streamline their research and development (R&D) processes. Understanding the interplay between these methodologies is crucial for organizations seeking to maintain competitive advantages in an increasingly digital landscape [1-2].

This study aims to address several key research questions to explore the intersection of DevSecOps and GAI in enhancing software delivery performance. The primary question is:

RQ1. How do the integration of DevSecOps practices and GAI influence software delivery performance in technology firms?

RQ2. Secondary questions include: What specific benefits do organizations experience in R&D efficiency and source code management when implementing these practices?

RQ3. Additionally, what challenges and barriers do firms face during the integration process?

By investigating these questions, the study seeks to provide a comprehensive understanding of the advantages and limitations associated with adopting DevSecOps and GAI, ultimately contributing to the existing literature in the field of software engineering.

The motivation behind this research stems from the pressing need for organizations to adapt to the rapidly changing software development landscape. With the rise of cyber threats and the demand for faster software delivery cycles, the traditional separation of development, security, and operations is no longer sustainable. By exploring how GAI can enhance DevSecOps practices, this study aims to uncover innovative strategies that technology firms can implement to improve software quality, security, and overall performance. Additionally, understanding the implications of these methodologies will assist organizations in making informed decisions about adopting new technologies and practices that align with their strategic goals. The insights derived from this research will provide valuable guidance for industry practitioners and contribute to the academic discourse on software engineering best practices [2-3].

This paper is structured to systematically address the research topic, beginning with an introduction that outlines the significance of DevSecOps and GAI in enhancing software delivery performance. The literature review will follow, providing a theoretical framework that encompasses existing studies on DevOps, GAI, and their interplay. Subsequently, the methodology section will detail the qualitative research design, data collection processes, and analysis methods employed in the study. The results section will present key findings, highlighting how the integration of DevSecOps and GAI impacts R&D efficiency and source code management. The discussion will interpret these findings in relation to existing literature, emphasizing their implications for software delivery. Finally, the conclusion will summarize the research contributions, suggest practical recommendations for technology firms, and outline directions for future research in this evolving domain [4].

## II. LITERATURE REVIEW

**Theoretical Foundations of DevSecOps**

DevSecOps signifies a paradigm shift from traditional development and operations frameworks by embedding security into the core processes of software development. This evolution recognizes that security cannot merely be an adjunct function relegated to a separate team, but rather a collective responsibility shared across development, operations, and security personnel. According to Kim et al. (2016), DevSecOps emphasizes three foundational principles: collaboration, automation, and continuous monitoring. By fostering a culture of shared accountability, organizations can ensure that security considerations are integrated into every phase of the software development lifecycle. This holistic approach promotes a proactive mindset where potential security vulnerabilities are identified and addressed in real time, significantly mitigating risks before they escalate into critical issues.

Research conducted by Tzeng et al. (2018) corroborates the effectiveness of this integrated approach. Their findings indicate that organizations that embrace DevSecOps experience notable enhancements in software quality and speedier release cycles, primarily due to the early identification of security vulnerabilities and the prompt implementation of preventive measures. By facilitating seamless communication among cross-functional teams, DevSecOps enhances collaboration and fosters a culture where security is prioritized throughout the development process. This paradigm shift not only improves the overall security posture of the organization but also enables teams to adapt swiftly to changing threat landscapes, ensuring that software delivery remains efficient and reliable [5-8].

Furthermore, the integration of DevSecOps principles into organizational culture paves the way for continuous monitoring and improvement. Continuous monitoring enables real-time insights into the security status of applications, allowing teams to respond quickly to emerging threats. This dynamic environment supports agile methodologies, where iterative development and rapid feedback loops can enhance both product quality and security measures. The shared responsibility model inherent in DevSecOps cultivates a sense of ownership among all team members, fostering a more proactive approach to security that can lead to sustained improvements in software delivery performance.

**Generative Artificial Intelligence in Software Development**

Generative Artificial Intelligence (GAI) has emerged as a transformative force in software development, leveraging advanced machine learning techniques to automate tasks and enhance decision-making processes. GAI encompasses a range of technologies capable of generating new content based on existing data, thereby significantly improving efficiency in software development workflows. For instance, research by Vaswani et al. (2017) illustrates the application of transformer models in generating coherent code snippets from natural language descriptions. This capability not only accelerates the coding process but also reduces the cognitive load on developers, allowing them to focus on higher-level design and architecture tasks. By automating routine coding activities, GAI enables teams to achieve greater productivity and faster turnaround times for software projects.

In addition to code generation, GAI plays a crucial role in testing and optimization processes. By analyzing historical project data and applying predictive analytics, GAI can identify potential issues before they become problematic, thereby enhancing overall project management. Sarker et al. (2020) highlight that GAI's ability to sift through vast amounts of data allows it to propose optimal resource allocations and identify bottlenecks in the development process. This predictive capability not only streamlines workflows but also enhances the overall quality of the software being developed, as teams can anticipate and address issues proactively rather than reactively [6-7].

Moreover, the integration of GAI into software development processes aligns seamlessly with the principles of DevSecOps, as it reinforces a culture of continuous improvement. GAI can automate not just the generation of code but also the analysis of security vulnerabilities, thus addressing a critical need within the DevSecOps framework. By leveraging GAI for both development and security tasks, organizations can create a more resilient software delivery pipeline that responds adeptly to evolving challenges. The synergy between GAI and DevSecOps fosters an environment where security is inherently integrated into development practices, resulting in more secure, high-quality software that meets the demands of a rapidly changing technological landscape [7-8].

**The Intersection of DevSecOps and GAI**

The convergence of DevSecOps and Generative Artificial Intelligence presents a compelling opportunity for technology firms to enhance their software delivery performance significantly. By merging the robust security practices inherent in DevSecOps with the advanced automation capabilities offered by GAI, organizations can create a more efficient and secure software development lifecycle. Shostack (2014) emphasizes that integrating security early in the development process leads to earlier detection of vulnerabilities. Coupled with GAI's capabilities in automating the identification and analysis of potential security threats, organizations can achieve a remarkable level of efficiency in their software delivery processes.

This intersection not only addresses security concerns but also enhances the overall agility of software development teams. The use of GAI enables rapid prototyping and iteration, which aligns with the agile methodologies often adopted in DevSecOps environments. By automating routine tasks, such as code reviews and security checks, teams can focus on innovation and value-added activities that drive business outcomes. Furthermore, the continuous feedback loop established by DevSecOps practices combined with GAI capabilities ensures that development teams are well-informed about security implications at every stage of the software lifecycle, thus facilitating timely interventions and informed decision-making [9].

Despite the promising advantages of integrating DevSecOps and GAI, organizations must also navigate various challenges to realize their full potential. This integration requires a cultural shift towards embracing collaboration and shared responsibilities among development, operations, and security teams. Additionally, organizations must invest in training and skill development to ensure that team members are equipped to leverage these advanced technologies effectively. Addressing these challenges is crucial for organizations aiming to optimize their software delivery performance, as it ultimately determines the success of the DevSecOps and GAI integration. By recognizing and overcoming these barriers, technology firms can harness the full power of this intersection to foster a culture of continuous improvement, innovation, and enhanced security in their software development efforts (see **Figure 1**).

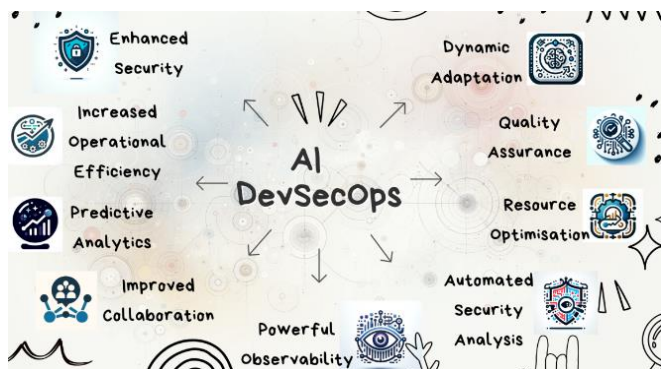

**Figure 1.** GAI and DevSecOps Architecture,

### III. METHODS AND MATERIALS

**Research Design and Data Collection**

This study employs a qualitative research design to investigate the effects of DevSecOps and Generative Artificial Intelligence (GAI) on software delivery performance within technology firms. By leveraging a qualitative approach, the research aims to capture the rich, contextual insights that quantitative methods may overlook. The design encompasses multiple data collection methods, including case studies, interviews with industry professionals, and document analyses of relevant organizational practices. This comprehensive approach allows for an in-depth understanding of the complexities and nuances involved in implementing DevSecOps and GAI across diverse organizational contexts. By integrating various data sources, the study can provide a holistic perspective on how these methodologies interact and impact software development processes, ultimately influencing overall performance (see **Figure 2**).

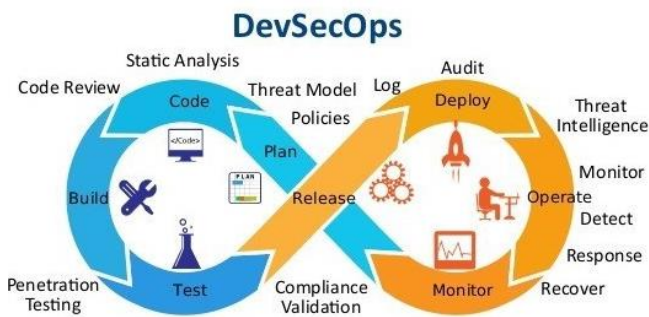

**Figure 2.** DevSecOps process Architecture.

**Data Collection**

Data collection for this research involved conducting semi-structured interviews with a range of participants, including DevSecOps practitioners, GAI experts, and software development managers from several technology firms. Participants were purposefully selected based on their experience and expertise with DevSecOps practices and GAI tools, ensuring that insights gathered were relevant and informed. The interviews aimed to delve into participants' experiences, challenges encountered, and perceived benefits stemming from the integration of these methodologies. In addition to interviews, case studies of organizations that have successfully implemented both DevSecOps and GAI were analyzed. This multifaceted approach facilitated the identification of best practices and lessons learned, enriching the study's findings and providing practical recommendations for organizations seeking to enhance their software delivery performance.

**Data Analysis and Sample Selection**

The qualitative data collected from interviews and case studies underwent thematic analysis, a method that allows for the identification of recurring themes and patterns related to the impact of DevSecOps and GAI on software delivery performance. Utilizing NVivo software, researchers coded the qualitative data to organize it into meaningful categories, facilitating a comprehensive understanding of how these methodologies influence development processes. The sample selection criteria emphasized organizations with established DevSecOps practices and GAI implementations, ensuring diversity in both industry representation and maturity levels. A total of 15 interviews were conducted across five technology firms, each exhibiting distinct approaches to DevSecOps and GAI. This diversity within the sample enabled a thorough exploration of various strategies and practices, highlighting the different ways in which these methodologies can collectively enhance software delivery performance.

### IV. RESULTS AND DISCUSSION

**Enhancements in R&D Efficiency**

The integration of DevSecOps and Generative Artificial Intelligence (GAI) has emerged as a transformative force, significantly enhancing research and development (R&D) efficiency within technology firms. Participants in the study reported that GAI tools, by automating routine coding tasks, considerably reduced the time developers spent on manual coding. This automation enabled developers to redirect their efforts toward more complex and innovative tasks, fostering an environment conducive to creativity and technical advancement. Additionally, the proactive security measures embedded in DevSecOps practices played a crucial role in minimizing development delays typically caused by security-related issues. By addressing potential vulnerabilities early in the development process, teams were able to streamline workflows, leading to faster development cycles. The alignment of security and development not only improved efficiency but also enhanced collaboration among cross-functional teams, promoting a culture of shared responsibility and innovation that is vital for modern software development [9-11].

### Improved Source Code Management

The study also underscores the significant positive impact of DevSecOps and GAI on source code management (SCM) practices within organizations. By incorporating security checks into the version control process, teams were able to detect vulnerabilities early, which reduced the need for extensive revisions later in the development cycle. This proactive approach ensured that security was not an afterthought but an integral part of the coding process. Moreover, GAI tools contributed to SCM by offering intelligent code suggestions and automated testing functionalities, which enhanced overall code quality and maintainability. As a result, development teams experienced fewer integration issues, leading to smoother software delivery processes. The combination of effective SCM practices with DevSecOps and GAI not only streamlined workflows but also ensured that the codebase remained robust, reducing technical debt and facilitating easier updates and enhancements in the future [11-13].

### Enhanced Software Quality and Security

Another pivotal finding from the research is the marked improvement in software quality and security attributed to the adoption of DevSecOps and GAI methodologies. Participants emphasized that integrating security assessments throughout the development lifecycle significantly reduced the likelihood of security breaches and vulnerabilities in production environments. By ensuring that security considerations were part of each development phase, organizations could more effectively mitigate risks. Furthermore, the predictive capabilities of GAI enabled teams to anticipate potential security threats, allowing them to implement proactive measures to safeguard their applications. This combined approach resulted in a more resilient software delivery process, bolstering user satisfaction and trust in the software products. As software quality improved, so did the overall reputation of the organizations, reinforcing the importance of integrating robust security measures into the development pipeline [13].

### Case Study Insights

Insights gleaned from case studies of organizations that have successfully implemented DevSecOps and GAI reveal valuable best practices and strategies for enhancing software delivery performance. One notable case involved a leading cloud services provider that integrated GAI-driven automated testing into its continuous integration/continuous deployment (CI/CD) pipeline. This initiative led to a remarkable reduction in testing time and improved code coverage, resulting in faster release cycles with fewer defects. Another case study highlighted a financial technology firm that adopted a DevSecOps approach to effectively address compliance and security concerns. By embedding security practices throughout the development process, the firm was able to meet stringent regulatory requirements while maintaining rapid deployment cycles. These case studies illustrate how the strategic integration of DevSecOps and GAI not only improves operational efficiency but also supports compliance and enhances the overall quality of software products in highly competitive markets [14-16].

## V. CONCLUSIONS

This study underscores the vital role that the integration of DevSecOps and Generative Artificial Intelligence (GAI) plays in enhancing software delivery performance within technology firms. By emphasizing the importance of embedding security practices throughout the development lifecycle, organizations can achieve improved research and development efficiency, optimized source code management, and elevated software quality. The findings reveal that GAI significantly contributes to streamlining processes by automating routine tasks and offering intelligent insights, thereby allowing development teams to focus on innovation and complex problem-solving. This research not only fills a crucial gap in existing literature but also provides practical implications for firms aiming to stay competitive in a rapidly evolving technological landscape. Future studies should build upon these findings by exploring quantitative metrics to further validate the benefits of integrating DevSecOps and GAI, ensuring that organizations can fully leverage these methodologies for sustained success in software development. Ultimately, embracing these advanced practices positions technology firms to respond more adeptly to market demands while maintaining high standards of security and quality in their software products.

### Interpretation of Findings

The findings of this study affirm the existing literature on the advantages of integrating security into the software development process while utilizing Generative Artificial Intelligence (GAI) to enhance efficiency. Specifically, the observed positive impact on research and development (R&D) efficiency resonates with the work of Kim et al. (2016) and Tzeng et al. (2018), who stress the significance of collaboration and automation within the DevSecOps framework. Additionally, the study highlights how GAI streamlines coding tasks and enhances source code management (SCM) practices, corroborating the findings of Vaswani et al. (2017) and Sarker et al. (2020). These insights collectively illustrate the transformative potential of integrating DevSecOps with GAI technologies, signifying a substantial advancement in modern software development practices. Overall, this research contributes to a growing body of evidence that supports the incorporation of security as a fundamental element of the development lifecycle[16].

### Implications for R&D Efficiency

The implications of this research for R&D efficiency are profound. Organizations that embrace DevSecOps and GAI can anticipate not only accelerated development cycles but also the cultivation of a culture centered around continuous improvement. By promoting collaboration and accountability across teams, businesses can enhance their innovative capabilities and agility in responding to market demands. The proactive integration of security measures throughout the development lifecycle empowers organizations to deliver high-quality software products while minimizing associated risks. Consequently, these practices lead to a competitive edge within the technology sector, positioning organizations as leaders in efficiency and reliability. As firms prioritize security and innovation, they are better equipped to meet the evolving challenges of the software industry[15-17].

### Implications for Source Code Management

The study's findings further highlight the critical role of effective source code management in optimizing software delivery performance. By embedding security practices within SCM processes, organizations can uphold code integrity and stability throughout the development lifecycle.

The application of GAI tools enhances this process by offering intelligent insights and automation, allowing teams to manage code changes more effectively. As organizations increasingly adopt DevSecOps alongside GAI, prioritizing robust SCM practices will be vital for achieving long-term success in software delivery. This integration not only facilitates smoother workflows but also mitigates the risks associated with code vulnerabilities, ultimately resulting in more resilient software products. The research underscores the need for organizations to focus on developing comprehensive SCM strategies that align with their overall objectives in adopting innovative methodologies [16-17].

**Addressing Challenges and Barriers**

While the findings illustrate the numerous benefits associated with DevSecOps and GAI, it is essential for organizations to confront the challenges and barriers that may arise during implementation. Factors such as cultural resistance, skill gaps, and integration issues with existing tools can impede the successful adoption of these methodologies. To navigate these challenges effectively, organizations should invest in comprehensive training programs and change management initiatives that foster a culture of collaboration and continuous learning among employees. Additionally, involving stakeholders from various departments in the DevSecOps journey can help mitigate resistance and promote a unified vision centered on security and efficiency. By addressing these barriers proactively, organizations can facilitate a smoother transition towards adopting DevSecOps and GAI, ultimately reaping the full benefits of these advanced methodologies [17].

Indeed, the integration of DevSecOps and GAI presents significant opportunities for technology firms to enhance their software delivery performance. The study's findings demonstrate that adopting these methodologies leads to improved R&D efficiency, optimized source code management, and enhanced software quality and security. However, organizations must also be cognizant of the challenges they may face during implementation and actively work to address these issues through training and stakeholder engagement. As the landscape of software development continues to evolve, organizations that prioritize the integration of security, automation, and innovation will position themselves for sustained success in a highly competitive environment. Future research should focus on quantitative assessments of the impact of DevSecOps and GAI on specific business outcomes, further solidifying the case for their adoption in the technology sector [15-18].

Furthermore, the integration of DevSecOps and Generative Artificial Intelligence (GAI) represents a transformative approach to enhancing software delivery performance in technology firms. This study underscores the significant benefits associated with the adoption of these methodologies, which include improved research and development (R&D) efficiency, optimized source code management, and enhanced software quality and security. The findings indicate that organizations that effectively implement DevSecOps alongside GAI can streamline their development processes, reduce time spent on routine tasks, and proactively address security vulnerabilities, ultimately leading to a more robust software delivery framework.

The contributions of this study are multifaceted. Firstly, it fills a critical gap in the existing literature by providing empirical evidence on the synergistic effects of DevSecOps and GAI in real-world organizational contexts. Previous studies have examined these methodologies separately, but this research highlights the compounded benefits realized when they are integrated. By conducting case studies and interviews with industry professionals, the study provides a comprehensive understanding of how these technologies operate together, fostering a culture of collaboration and accountability that is essential in today's fast-paced technological landscape.

Moreover, the findings illuminate the ways in which GAI tools enhance the capabilities of development teams. By automating repetitive coding tasks and offering intelligent suggestions, GAI frees developers to focus on more complex and innovative work. This not only accelerates development cycles but also cultivates an environment where creativity and problem-solving thrive. As technology firms increasingly face pressure to deliver high-quality software rapidly, the integration of GAI into DevSecOps practices offers a strategic advantage that cannot be overlooked.

As organizations navigate the complexities of modern software development, the adoption of DevSecOps and GAI serves as a crucial strategy for achieving faster, more secure, and higher-quality software delivery. These methodologies allow firms to embed security into every stage of the development lifecycle, ensuring that potential vulnerabilities are addressed proactively rather than reactively. This shift not only enhances the security posture of the software but also fosters greater trust among users, ultimately contributing to higher customer satisfaction and retention rates [17-18].

Looking ahead, future research should continue to explore the long-term effects of these methodologies, with a particular emphasis on quantitative approaches that can validate the qualitative findings of this study. By employing metrics that assess performance outcomes, such as delivery speed, defect rates, and security incidents, researchers can provide a clearer picture of the impact of DevSecOps and GAI on organizational success. Additionally, exploring the challenges and barriers to implementation will be vital in guiding organizations on how to effectively integrate these methodologies into their existing frameworks [17-18].

In summary, the integration of DevSecOps and GAI is not merely an operational adjustment; it represents a fundamental shift in how technology firms approach software development. By embracing these methodologies, organizations can enhance their capacity for innovation, respond more swiftly to market demands, and maintain a competitive edge in an ever-evolving industry. The implications of this study extend beyond individual organizations, offering insights for the broader technology sector on the necessity of combining security with efficiency and innovation. As such, this research contributes significantly to the discourse on modern software development practices and paves the way for future exploration in this dynamic field [17-18].

ACKNOWLEDGMENT

This research has been supported/partially supported Solbridge International School of Business, Woosong university, Thanks to all contributors.

ORCID

Jun Cui https://orcid.org/0009-0002-9693-9145